\documentclass{article}

\usepackage{microtype}
\usepackage{graphicx}
\usepackage{subfigure}
\usepackage{booktabs} 
\usepackage{url}
\usepackage{amsmath,amssymb,amsfonts}
\usepackage{amsthm}
\usepackage{thmtools}
\usepackage{float}
\usepackage{pifont}
\newcommand{\indep}{\perp \!\!\! \perp}
\usepackage{amsthm}

\usepackage{xcolor}
\definecolor{expert}{HTML}{008000}
\definecolor{error}{HTML}{f96565}

\usepackage{hyperref}
\hypersetup{
  colorlinks=true,
  linkcolor=blue!50!red,
  urlcolor=green!70!black
}

\usepackage[accepted]{icml2021}

\usepackage{tikz}
\usetikzlibrary{arrows,calc,bayesnet} 
\newcommand{\tikzAngleOfLine}{\tikz@AngleOfLine}
\def\tikz@AngleOfLine(#1)(#2)#3{%
\pgfmathanglebetweenpoints{%
\pgfpointanchor{#1}{center}}{%
\pgfpointanchor{#2}{center}}
\pgfmathsetmacro{#3}{\pgfmathresult}%
}

\declaretheoremstyle[
headfont=\normalfont\itshape,
qed=\qedsymbol,
]{mypf}
\declaretheorem[numbered=no, name=Proof, style=mypf]{pf}

\newcommand{\sref}[1]{Sec. \ref{#1}}

\icmltitlerunning{Game-Theoretic Algorithms for Conditional Moment Matching}

\begin{document}

\twocolumn[
\icmltitle{Game-Theoretic Algorithms for Conditional Moment Matching}



\icmlsetsymbol{equal}{*}

\begin{icmlauthorlist}
\icmlauthor{Gokul Swamy}{ri}
\icmlauthor{Sanjiban Choudhury}{aurora}
\icmlauthor{J. Andrew Bagnell}{aurora,ri}
\icmlauthor{Zhiwei Steven Wu}{isr}
\end{icmlauthorlist}

\icmlaffiliation{ri}{Robotics Institute, Carnegie Mellon University}
\icmlaffiliation{isr}{Institute for Software Research, Carnegie Mellon University}
\icmlaffiliation{aurora}{Aurora Innovation}

\icmlcorrespondingauthor{Gokul Swamy}{gswamy@cmu.edu}

\icmlkeywords{Machine Learning, ICML}

\vskip 0.3in
]



\printAffiliationsAndNotice{} 

\begin{abstract}
A variety of problems in econometrics and machine learning, including instrumental variable regression and Bellman residual minimization, can be formulated as satisfying a set of conditional moment restrictions (CMR). We derive a general, game-theoretic strategy for satisfying CMR that scales to nonlinear problems, is amenable to gradient-based optimization, and is able to account for finite sample uncertainty. We recover the approaches of \cite{dikkala2020minimax} and \cite{dai2018sbeed} as special cases of our general framework before detailing various extensions and how to efficiently solve the game defined by CMR.
\end{abstract}

\section{Introduction}
Let $X$, $Y$, and $Z$ be random variables on (potentially non-finite) sample spaces $\mathcal{X}$, $\mathcal{Y}$, and $\mathcal{Z}$. We are interested in finding a function $h \in \mathcal{H}$ that satisfies a set of \textit{conditional moment restrictions} \cite{chamberlain1987asymptotic},
\begin{equation}
\label{eq:cmr}
    \forall z \in \mathcal{Z},\, \mathbb{E}[Y|z] = \mathbb{E}[h(X)|z],
\end{equation}
or CMR for short. While this problem might seem a bit abstract, it is at the core of two disparate problems in machine learning: \textit{instrumental variable regression} and \textit{Bellman-residual minimization}. We give a brief introduction to each before presenting a unified method for solving for consistent $h$ from finite samples that elegantly scales to nonlinear problems. Throughout, we assume we optimize over a class $\mathcal{H}$ that is convex, compact, closed under negation, and of finite Rademacher complexity.

Our key insight is that \textit{we can formulate conditional moment matching as a zero-sum game, allowing us to both eliminate double sample issues and explicitly reason about the effects of constraint relaxation}. We call this family of techniques \textit{\textbf{estimation via relaxation}}.

\subsection{Instrumental Variable Regression}
Let's assume that $X$, $Y$, and $Z$ have the following dependency structure:
\begin{figure}[h]
    \centering
\begin{tikzpicture}[scale=1, transform shape]
    \node (a) [draw, very thick, circle, fill=lightgray] at (0.0, 0) {$Z$};
    \node (b) [draw, very thick, circle, fill=lightgray] at (1.5, 0) {$X$};
    \node (c) [draw, very thick, circle, fill=lightgray]  at (3, 0) {$Y$};
    \node (d) [draw, very thick, circle]  at (1.5, -1.5) {$U$};
    \path[->, color=black] (a) to[bend right] node[midway] {$g$} (b);
    \draw [->, very thick] (a) to (b);
    \path[->, color=expert] (b) to[bend right] node[midway] {$h$} (c);
    \draw [->, very thick, color=expert] (b) to (c);
    \draw [->, very thick] (d) to (b);
    \draw [->, very thick] (d) to (c);
    \end{tikzpicture}
    \caption{The graphical model considered in instrumental variable regression. We are interested in finding $h$, the causal relationship from $X$ to $Y$, even though there is an unobserved confounder, $U$.}
\end{figure}
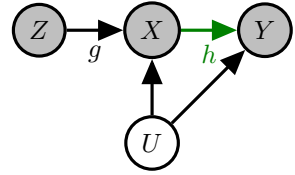

We refer to $X$ as the \textit{treatment} and $Y$ as the \textit{response} or \textit{outcome}. Given a dataset of $(x, y, z)$ tuples, we are interested in determining the causal relationship between $X$ and $Y$, $\mathbb{E}[Y|\text{do}(x)]$. In the above graphical model, this is equivalent to determining $h$. Because of the presence of an unobserved confounder, $U$, that affects both $X$ and $Y$, standard regression (e.g. Ordinary Least Squares or OLS) can produce inconsistent estimates. If we only have observational data and are unable to perform randomized control trials, a canonical technique to recover $h$ is Instrumental Variable Regression (IVR) \cite{winship1999estimation}. Formally, an instrument $Z$ must satisfy:
\vspace{-4pt}
\begin{enumerate}
    \item \textit{Unconfounded Instrument}: $Z \indep U$ -- i.e. ``independent" randomization from instrument. \vspace{-4pt}
    \item \textit{Exclusion}: $Z \indep Y | X$ -- i.e. no extraneous paths. \vspace{-4pt}
    \item \textit{Relevance}: $Z \not\!\perp\!\!\!\perp X$ -- i.e. conditioning has an effect. \vspace{-4pt}
\end{enumerate}
\vspace{-4pt}
We note that the unconfounded instrument and exclusion conditions are structural assumptions that cannot be checked from observational data. Assuming access to such a $Z$ and linear relationships between all variables, one can recover $h(x) = \beta x$ by computing $\beta = \mathbb{E}[ZX]^{-1}\mathbb{E}[ZY]$. Notice that the recovered $h$ satisfies the CMR \eqref{eq:cmr}. With a bit of linear algebra, one can show this calculation is equivalent to a Two-Stage Ordinary Least Squares (2SLS) procedure, for which one first regresses from $Z$ to $X$ and then regresses from the predicted $\hat{X}$ to $Y$, returning the latter coefficients.

We focus on the general, nonlinear problem, which can still be formulated in terms of CMRs. We assume that noise $U$ enters additively to $Y$ \footnote{This is an assumption inherited from standard regression. Without it, one can only bound the treatment effect \cite{kilbertus2020class}.}, and write out the following equations:
\begin{equation}
    X = g(Z, U),
    \vspace{-4pt}
\end{equation}
\begin{equation}
    Y = h(X) + U.
    \vspace{-4pt}
\end{equation}

Without loss of generality, we assume that $\mathbb{E}[U] = 0$. As is standard in the  causal inference literature \cite{nie2020learning}, we assume that our collected data satisfies consistency of potential outcomes (i.e. the data generating process did not change since our dataset was collected \cite{rehkopf2016consistency}) and overlap (i.e. we have data for all $X$ we care about \cite{overlap}). Perhaps the most natural way to tackle this problem would be to minimize some notion of squared error between the two sides of the CMR:
\begin{equation}
\label{eq:squared}
    \min_{h \in \mathcal{H}} \mathbb{E}_z[\mathbb{E}[Y-h(x)|z]^2].
    \vspace{-4pt}
\end{equation}
For nonlinear problems, gradient-based optimization is a common and scalable technique. Unfortunately, differentiating the squared expectation in the preceding expression leads to a ``double sample" issue in which one needs two independent samples from $P(X|z)$ to compute each gradient step, which can be quite challenging outside of simulated environments. To resolve this issue, \cite{dikkala2020minimax} propose instead solving the following zero-sum game:
\begin{equation}
\label{eq:dikkala}
    \min_{h \in \mathcal{H}} \max_{f}  \mathbb{E}[2(Y - h(X))f(Z) - f(Z)^2],
\end{equation}
which they prove produces consistent estimates of $h$ and does not involve a squared expectation. We show in \sref{sec:emr} that this is an example of a general technique for satisfying CMR.

\subsection{Bellman Residual Minimization}
Consider a Markov Decision Process (MDP) parameterized by $\langle \mathcal{S}, \mathcal{A}, \mathcal{T}, r, \gamma \rangle$, where $\mathcal{S}$ is the state space, $\mathcal{A}$ is the action space, $\mathcal{T}: \mathcal{S} \times \mathcal{A} \rightarrow \Delta(\mathcal{S})$ is the transition operator, $r: \mathcal{S} \times \mathcal{A} \rightarrow [-1, 1]$ is a reward function, and $\gamma$ is the discount factor. It is a well-known fact that one can find the optimal policy, $\pi$, for this MDP by first computing the optimal \textit{value function} via the Bellman Equation,
\begin{equation}
\label{eq:bellman}
    V(s) = \max_{a \in \mathcal{A}} r(s, a) + \gamma \mathbb{E}_{s' \sim \mathcal{T}(s, a)}[V(s')],
\end{equation}
and choosing actions greedily \cite{puterman2014markov}:
\begin{equation}
\label{eq:greedy}
    \pi(s) = \arg\max_{a \in \mathcal{A}} r(s, a) + \gamma \mathbb{E}_{s' \sim \mathcal{T}(s, a)}[V(s')].
\end{equation}
We can combine these two expressions into one that is satisfied by the optimal $(V, \pi)$ pair:
\begin{equation}
\label{eq:comb}
    V(s) =  \mathbb{E}_{\substack{a \sim \pi(s) \\s' \sim \mathcal{T}(s, a)}}[r(s, a) + \gamma V(s')].
\end{equation}
When one cannot simply enumerate all states and actions to perform either policy or value iteration \cite{russell2002artificial}, a standard technique is \textit{Bellman residual minimization} -- minimizing the difference between the two sides of \eqref{eq:comb}. The updated value function can then be used to compute a more optimal policy via the greedy optimization of \eqref{eq:greedy}. We can write \eqref{eq:comb} in the template of CMR by setting
\begin{equation}
    Z = s, a
    \vspace{-4pt}
\end{equation}
\begin{equation}
    X = s,\, h(X) = V(s)
    \vspace{-4pt}
\end{equation}
\begin{equation}
    Y =  r(s, a) + \gamma \mathbb{E}_{s' \sim \mathcal{T}(s, a)}[V(s')].
    \vspace{-4pt}
\end{equation}
Similar to \eqref{eq:squared}, one could attempt to satisfy the CMR by minimizing the expected squared Bellman error:
\begin{equation}
    \min_{V: \mathcal{S} \rightarrow \mathbb{R}} \mathbb{E}[\mathbb{E}[V(s) - (r(s, a) + \gamma \mathbb{E}_{s' \sim \mathcal{T}(s, a)}[V(s')])|s, a]^2],
    \vspace{-4pt}
\end{equation}
which also has double sample issues \cite{baird1995residual}. Recently, \cite{dai2018sbeed} suggested that one could avoid double sample issues by instead solving the following zero-sum game:\footnote{For simplicity, we drop the entropy regularization in their objective. Our derivations can be easily extended.}
\begin{equation}
\label{eq:lihong}
    \min_{V: \mathcal{S} \rightarrow \mathbb{R}} \max_{f: \mathcal{S} \times \mathcal{A} \rightarrow \mathbb{R}} \mathbb{E}_{s, a, s'}[2(r(s, a) + V(s') - V(s))f(s, a) - f(s, a)^2],
    \vspace{-4pt}
\end{equation}
which they derive via an appeal to convex conjugates \cite{boyd2004convex}. We present an alternative construction of this objective which helps elucidate the properties of the algorithm proposed by \cite{dai2018sbeed}.

\section{Estimation via Relaxation}
\begin{table}[h]
\label{cmrtable}
\vskip 0.15in
\begin{center}
\begin{small}
\begin{sc}
\begin{tabular}{lcccr}
\toprule
 Setting & $Z$ & $X$ and $h(X)$ & $Y$\\
\midrule
 IVR & Instrument & Treatment & Outcome \\
 RL & $(s, a)$ & $s$ and $V(s)$ & $r(s, a) + \gamma \mathbb{E}_{s'}[V(s')]$ \\
\bottomrule
\end{tabular}
\end{sc}
\end{small}
\end{center}
\vskip -0.1in
\caption{The work of both \cite{dikkala2020minimax} and \cite{dai2018sbeed} can be seen as examples of a more general template for satisfying conditional moment restrictions from finite samples and for nonlinear problems.}
\end{table}
\label{sec:emr}
Throughout this section, we use \textbf{bold fonts} to designate vectors. To aid us in our quest to satisfy the conditional moment restrictions, we are given access to $N$ samples from the joint distribution of $X$, $Y$, and $Z$. Because we have finite samples and can therefore only estimate conditional expectations up to some tolerance, it is natural to relax the CMRs to
\begin{equation}
\label{eq:tik}
\begin{array}{ll@{}ll}
\min_{h \in \mathcal{H},\, \boldsymbol{\delta}}  &  J(\boldsymbol{\delta})         \\
\text{s.t.} & |\mathop{{}\mathbb{E}}[Y - h(X)|z]| \leq \delta_z \quad \forall z \in \mathcal{Z},\\
\end{array}
\end{equation}
where the $\delta_z$ are slack variables and $J$ is some convex function of $\boldsymbol{\delta}$ that keeps the slacks from getting too large. We note that \eqref{eq:tik} is reminiscent of the classical framework of Tikhinov regularization before presenting some interesting properties that result from setting $J(\boldsymbol{\delta}) = \frac{1}{2}\mathbb{E}_z[\delta_z^2]$. The Lagrangian with a $P(z)$-weighted inner product is
\begin{equation}
    L(h, \boldsymbol{\delta}, \boldsymbol{\lambda}) = \sum_{z \in \mathcal{Z}} P(z)\lambda_z(\mathbb{E}[Y - h(X)|z] - \delta_z) + P(z)\frac{1}{2}\delta_z^2,
\end{equation}
where $\boldsymbol{\lambda}$ are the Lagrange multipliers.\footnote{While written above in terms of finite $z$, our derivation easily extend to infinite sets.} Applying the stationarity component of the KKT conditions, one arrives at
\begin{equation}
    \nabla_{\delta_z} L(h, \boldsymbol{\delta}, \boldsymbol{\lambda}) = -P(z)\lambda_z + P(z)\delta_z = 0,
\end{equation}
implying that $\delta_z = \lambda_z$. Plugging this back into the Lagrangian, we can simplify our function to
\begin{equation}
\label{eq:rela}
    L(h, \boldsymbol{\lambda}) = \sum_{z \in \mathcal{Z}} P(z)\lambda_z\mathbb{E}[Y - h(X)|z] - P(z)\frac{1}{2}\lambda_z^2.
\end{equation}
We refer to \eqref{eq:rela} as the \textit{Regularized Lagrangian} or ReLa for
short. Now, solving for the optimal Lagrange multipliers via stationarity, we arrive at
\begin{equation}
    \nabla_{\lambda_z} L(h, \boldsymbol{\lambda}) = P(z)\mathbb{E}[Y - h(X)|z] - P(z)\lambda_z = 0,
\end{equation}
which implies the equilibrium $\lambda_z$ is equal to $\mathbb{E}[Y - h(X)|z]$. Plugging this back into \eqref{eq:rela} recovers function
\begin{equation}
\label{eq:chen}
    L(h) = \sum_{z \in \mathcal{Z}} P(z) \mathbb{E}[Y - h(X)|z]^2.
\end{equation}
Thus, in the population limit, we are optimizing the conditional MSE of \cite{chen2012estimation}, leading to consistent estimates. 
\subsection{Generative Modeling Approach}
Perhaps the most immediate way to minimize \eqref{eq:chen} over $h \in \mathcal{H}$ would be to minimize the empirical MSE,
\begin{equation}
    \min_{h \in \mathcal{H}} \frac{1}{N}\sum_i^N (y_i - h(x_i))^2.
\end{equation}
Unfortunately, this only gives us a function that matches unconditional moments (i.e. $\mathbb{E}[Y] = \mathbb{E}[h(X)]$). For the IVR setting, this would give us the inconsistent, OLS-like estimates of $h$ we are explicitly trying to avoid. For the MDP setting, this would not produce a valid value function.

 Instead, one could learn the distribution $P(X|z) = g(z)$ and pass samples from it to a candidate $h$, ensuring one is attempting to match the conditional moments. Intuitively, this is the generalization of the 2SLS procedure to nonlinear functions. Because the second stage is nonlinear, one cannot simply compute the first moment of the $P(X|z)$ distribution (which is recovered by linearly regressing from $X$ to $Z$ in the 2SLS procedure). To see this, consider $\mathcal{H}$ being the set of quadratic functions of $X$. Then,
 \begin{align}
    \mathbb{E}[h(X)|z] &= \sum_{x \in \mathcal{X}}p(x|z) h(x) \\
    & =\sum_{x \in \mathcal{X}}p(x|z) (ax^2 +bx + c) \\
    &= a\mathbb{E}[X^2|z] + b\mathbb{E}[X|z] + c.
 \end{align}
Here, one needs to have access to $\mathbb{E}[X^2|z]$ to check the CMR. For more complex $\mathcal{H}$, one therefore needs to learn the entire $P(X|z)$. This kind of approach was first proposed for the IVR setting by \cite{hartford17a} and amounts to first learning a $g(z)$ via maximum likelihood estimation and then solving
\begin{equation}
\label{eq:double_sample}
    \min_{h \in \mathcal{H}} \frac{1}{N}\sum_i^N (y_i - \mathbb{E}_{\hat{x} \sim g(z_i)}[h(\hat{x})])^2.
\end{equation}
Unfortunately, this approach suffers from the well-known ``double-sample" issue where multiple independent samples from $g(z)$ are required to compute gradients of $h$. To see this, note that the gradient w.r.t. $h$ of \eqref{eq:double_sample} is
\begin{equation}
    \sum_i^N (y_i - \mathbb{E}_{\hat{x} \sim \hat{g}(z_i)}[h(\hat{x})])(- \mathbb{E}_{\hat{x} \sim \hat{g}(z_i)}[\frac{\partial}{\partial h}h(\hat{x})]).
\end{equation}
Approximating this gradient with a single sample,
\begin{equation}
\sum_i^N (\mathop{{}\mathbb{E}}_{\hat{x} \sim \hat{g}(z_i)}[(y_i - h(\hat{x})) \frac{\partial}{\partial h}h(\hat{x})]),
\end{equation}
can produce biased estimates of the gradient that are inconsistent in the limit of infinite data. Additionally, learning an accurate first-stage model might be quite challenging for some problems.


\subsection{Game-Theoretic Approach}
Ideally, we would like to avoid the added complexity of learning a generative model and the double-sampling required for gradient based-optimization. One gets a two-for-one deal by instead solving the two-player zero-sum game with the ReLa \eqref{eq:rela} as the payoff. Denoting by $f \in \mathcal{F} \subseteq \{\mathcal{Z} \rightarrow \mathbb{R}\}$ the function that maps $z$'s to corresponding Lagrange multipliers, we can write this game as:
\begin{equation}
\label{eq:rela_game}
\min_{h \in \mathcal{H}} \max_{f \in \mathcal{F}}  \mathbb{E}[2(Y - h(X))f(Z) - f(Z)^2].
\end{equation}
Notice the similarity of this expression to the objectives of \cite{dikkala2020minimax} \eqref{eq:dikkala} and \cite{dai2018sbeed} \eqref{eq:lihong} and that there are no squared expectations of $h$ and therefore no double sample issues. Additionally, one does not need to learn a generative model of $P(X|z)$ for these sorts of game-theoretic approaches. Under the assumption that $\mathbb{E}[Y - h(X)|z] \in \mathcal{F}$, finding a Nash equilibrium of this game corresponds to finding a $h \in \mathcal{H}$ that is close in an $\ell_2$ sense to satisfying the CMR.

\subsection{Takeaways Thus Far}
We pause briefly to consider some of the important facts revealed to us by the above derivation:
\begin{enumerate}
    \item The approaches of \cite{dikkala2020minimax} and \cite{dai2018sbeed} are solving a relaxed CMR problem with a penalty on the $\ell_2$ norm of the constraint violation. Thus, their consistency is in an $\ell_2$ sense rather than a uniform, $\ell_{\infty}$ sense.
    \item The choice of $J$ allows us to control how slack is dispersed between the CMRs -- derivations like the above allow us to explicitly consider the effects of the kind of regularization we choose.
    \item The value of $f(z)$ (regardless of the choice of $J$) is exactly the amount of conditional moment mismatch the optimization procedure tolerates. If this value is particularly high for values of $z$ where we have many samples, we might need to consider a more expressive $\mathcal{H}$. Thus, we can explicitly reason about the effects of finite-sample uncertainty in satisfying CMRs.
\end{enumerate}



\section{Extensions}
We now briefly sketch two extensions to our above setup before diving into how to solve ReLa games efficiently.
\subsection{Constraints on $\boldsymbol{\delta}$}
A natural question after the above discussion might be how one could include information about sample uncertainty into the optimization procedure, rather than performing an after-the-fact check. Let $n_z$ denote the number of samples we have for $z$. For simplicity, we assume $n_z > 0,\, \forall z \in \mathcal{Z}$. A Hoeffding bound tells us that we should expect a sample expectation $\hat{\mathbb{E}}_{n_z}[\cdot|z]$ to be within $\propto \frac{1}{\sqrt{n_z}}$ of the population expectation. We refer to this sampling error as $\epsilon_z$. Then, in expectation over $Z$,
\begin{equation}
    \mathbb{E}_Z[\epsilon_z^2] = \sum_{z \in \mathcal{Z}}P(z) \epsilon_z^2 \approx \sum_{z \in \mathcal{Z}} \frac{n_z}{N} (\frac{1}{\sqrt{n_z}})^2 = \frac{|\mathcal{Z}|}{N} = \kappa(N).
\end{equation}
Our relaxed optimization problem then becomes
\begin{equation}
\label{eq:ivanov}
\begin{array}{ll@{}ll}
\min_{h \in \mathcal{H},\, \boldsymbol{\delta}}  & 0          \\
\text{s.t.} & |\mathop{{}\mathbb{E}}[Y - h(X)|z]| \leq \delta_z \quad \forall z \in \mathcal{Z}\\
& \mathbb{E}_Z[\delta_z^2] \leq \kappa(N).
\end{array}
\end{equation}
This form, reminiscent of Ivanov regularization, takes into account sample uncertainty explicitly and could be solved via standard constrained optimization machinery like Augmented Lagrangians \cite{hestenes1969multiplier}. We note that in practice, it is far more common to solve the Tikhanov-style problem \eqref{eq:tik} and scale $J$ based on performance on some holdout data.
\subsection{Regularization of $h$}
Consider two solutions $(h_1, f_1)$, $(h_2, f_2)$ of the ReLa Game \eqref{eq:rela_game} such that
\begin{align}
    &|\mathbb{E}[Y - h_1(X)|z]| \leq f_1(z),\, \forall z \in \mathcal{Z},\\
    &|\mathbb{E}[Y - h_2(X)|z]| \leq f_2(z),\, \forall z \in \mathcal{Z}, \\
    &\mathbb{E}_Z[f_1(z)^2] = \mathbb{E}_Z[f_2(z)^2].
\end{align}
Our game-theoretic perspective wouldn't be able to break ties between $h_1$ and $h_2$, regardless of desirable properties $h_1$ might have over $h_2$ like smoothness in $X$ or being the max-margin classifier. This is because as long as two solutions exist within CMR-violation balls of the same size (e.g. in $\ell_2$ norm), we consider them equally optimal. Thus, some sort of regularization on $h$, producing an optimization problem of the form
\begin{equation}
\label{eq:regh}
\begin{array}{ll@{}ll}
\min_{h \in \mathcal{H},\, \boldsymbol{\delta}}  &  J(\boldsymbol{\delta}) + \alpha R(h)       \\
\text{s.t.} & |\mathop{{}\mathbb{E}}[Y - h(X)|z]| \leq \delta_z \quad \forall z \in \mathcal{Z}\\
\end{array}
\end{equation}
with $\alpha$ tuned empirically on hold-out data might be helpful for finding $h$ with additional desirable properties. Concretely, one could regularize to the OLS solution by setting $R(h) = \mathbb{E}[(h(X) - Y)^2]$, encouraging $h$ to also match unconditional moments. Both \cite{dikkala2020minimax} and \cite{dai2018sbeed} consider regularization on $h$ (e.g. in $||\cdot||^2_{\mathcal{H}}$ norm or by entropy-regularizing the policy), which we identify as a form of tie-breaking.

\subsection{Efficiently Solving the ReLa Game}
\begin{algorithm}[t]
\begin{algorithmic}
\STATE {\bfseries Input:} Dataset $\mathcal{D}$ of $(x, y, z)$ tuples. No-regret algorithm over $\mathcal{H}$, Best-response oracle over $\mathcal{F}$, Threshold $\epsilon$ 
\STATE {\bfseries Output:} Causal effect of $X$ on $Y$, $h$
 \STATE Set $t = 1$, $h^t \in \mathcal{H}$, $f^t \in \mathcal{F}$, $L(h^t, f^t) = 2 \epsilon$
 \WHILE{$L(h^t, f^t) > \epsilon$}
 \STATE $L(h^t, f^t) = \mathop{{}\mathbb{E}}_{(x, y, z) \sim \mathcal{D}}[2(y - h^t(x))f^t(z) - f^t(z)^2]$
 \STATE No-regret alg. computes $h^t$ over $L(\cdot, f^t)$ history.
 \STATE Best-response computes $f^t = \arg\max_{f \in \mathcal{F}} L(h^t, \cdot)$.
 \STATE $t \leftarrow t + 1$
 \ENDWHILE
 \STATE Return $h^t$.
\end{algorithmic}
\caption{No-Regret Conditional Moment Matching}
\label{alg:cmr_nr}
\end{algorithm}
Solving a two-player game like \eqref{eq:rela_game} can be done provably efficiently via a reduction to no-regret online learning, following the classic analysis of \cite{freund1997decision}. For completeness, we provide such a procedure in Algorithm \ref{alg:cmr_nr} and prove efficiency below.
\begin{pf}
Let $L(h, f) =  \mathbb{E}[2(Y - h(X))f(Z) - f(Z)^2]$. Our no-regret assumption tells us that
\begin{equation}
    \frac{1}{N} \sum_t^N L(h^t, f^t) - \frac{1}{N} \min_{h \in \mathcal{H}} \sum_t^N L(h, f^t) \leq  \epsilon.
\end{equation}
for some $N$ that is $\text{poly}(\frac{1}{\epsilon})$. Under a realizability assumption, we can write that
\begin{equation}
   \frac{1}{N} \sum_t^N L(h^t, f^t) \leq \epsilon + \frac{1}{N} \min_{h \in \mathcal{H}} \sum_t^N L(h, f^t) \leq \epsilon.
\end{equation}
Then, utilizing the fact that there must be at least one element in an average that is at most the value of the average,
\begin{equation}
    \min_t L(h^t, f^t) \leq \frac{1}{N} \sum_t^N L(h^t, f^t) \leq \epsilon.
\end{equation}
To complete the proof, we recall that $f^t$ is chosen as the best response to $h^t$ in Algorithm \ref{alg:cmr_nr}, giving us that:
\begin{equation}
   \min_t \max_{f} L(h^t, f) \leq \epsilon
\end{equation}
Thus, the $h^t$ that minimizes $L(h^t, f^t)$ is half of an $\epsilon$-approximate Nash equilibrium of the ReLa game and can be computed within $\text{poly}(\frac{1}{\epsilon})$ iterations. In our setting, $\epsilon$ corresponds to the additional expected CMR violation our recovered $h$ suffers on top of the best element in $\mathcal{H}$.
\end{pf}
We note this reduction allows one to plug in \textit{any} no-regret algorithm (e.g. Follow the Regularized Leader \cite{37013} or Multiplicative Weights \cite{arora2012multiplicative}) and the efficiency and approximate CMR satisfaction result to still hold. Practically, one could instantiate this no-regret reduction via a GAN-like optimization procedure with the learning rate for the $h$ player much lower than that of the $f$ player, simulating the no-regret vs. best response iterations of Algorithm \ref{alg:cmr_nr}. While the connection to no-regret online learning is not made explicit, both \cite{dikkala2020minimax} and \cite{dai2018sbeed} follow an approach in this vein to scale their methods to high-dimensional tasks.

\section{Discussion}
We unify the techniques of \cite{dikkala2020minimax} and \cite{dai2018sbeed} under the umbrella of satisfying relaxed conditional moment restrictions, providing a framework that allows one to reason about sample uncertainty and the effects of regularization on such problems. We further consider explicit constraints on slack variables, the tie-breaking benefits of regularizing $h$, and provide a reduction to no-regret online learning for efficient solving of Regularized Lagrangian Games.

Moving forward, we would be interested in passing the higher bar of matching not just the first moments of $Y|z$ but the whole distribution, as well as fitting other problems into the framework of conditional moment restrictions. 

\section{Acknowledgements}
We thank Keegan Harris and Logan Stapleton for comments on a draft of this work. ZSW was supported in part by the NSF FAI Award \#1939606, a Google Faculty Research Award, a J.P. Morgan Faculty Award, a Facebook Research Award, and a Mozilla Research Grant. GS is supported by NSF Award \#1952085 and his family and friends.

\bibliography{example_paper}
\bibliographystyle{icml2021}



\end{document}